\documentclass{ifacconf}
\usepackage{amsmath,amssymb,amsfonts,amsbsy}
\usepackage{graphicx}
\usepackage{upgreek}
    
\usepackage{float}
\usepackage[utf8]{inputenc}
\usepackage{svg}
\usepackage{oubraces}
\usepackage{parskip}
\usepackage{natbib}        

\usepackage{eso-pic}

\AddToShipoutPictureBG*{%
	\AtPageUpperLeft{%
		\setlength\unitlength{1in}%
		\hspace*{\dimexpr0.5\paperwidth\relax}
		\makebox(0,-2.2)[c]{\begin{tabular}{c c}
		        Jilles van Hulst, Feedforward Control in the Presence of Input Nonlinearities: A Learning-based Approach,\\
				To appear at {\em Modeling, Estimation and Control Conference}, Jersey City, New Jersey, 2022.
		\end{tabular}}
}}

\AddToShipoutPictureBG*{%
	\AtPageUpperLeft{%
		\setlength\unitlength{1in}%
		\hspace*{\dimexpr0.5\paperwidth\relax}
		\makebox(0,-23.5)[c]{\begin{tabular}{c c}
				\small \copyright\ 2022 the authors. This work has been accepted to IFAC for publication under a Creative Commons Licence CC-BY-NC-ND.
		\end{tabular}}
}}

\begin{document}
\begin{frontmatter}

\title{Feedforward Control in the Presence of Input Nonlinearities: \linebreak A Learning-based Approach}

\thanks[footnoteinfo]{The research is supported by ASM Pacific Technology, Beuningen, The
Netherlands.}

\author[First]{Jilles van Hulst},
\author[First]{Maurice Poot},
\author[Second]{Dragan Kosti\a'c},
\author[Second]{Kai Wa Yan},
\author[Third]{Jim Portegies},
\author[First,Fourth]{Tom Oomen}

\address[First]{Control Systems Technology Section, Dept. of Mechanical Engineering, Eindhoven University of Technology, The Netherlands (e-mail: j.s.v.hulst@tue.nl).}
\address[Second]{Center of Competency, ASM Pacific Technology, The Netherlands}
\address[Third]{CASA, Dept. of Mathematics and Computer Science, Eindhoven University of Technology, The Netherlands}
\address[Fourth]{Delft Center for Systems and Control, Delft University of Technology, The Netherlands}

\begin{abstract}
Advanced feedforward control methods enable mechatronic systems to perform varying motion tasks with extreme accuracy and throughput. The aim of this paper is to develop a data-driven feedforward controller that addresses input nonlinearities, which are common in typical applications such as semiconductor back-end equipment. The developed method consists of parametric inverse-model feedforward that is optimized for tracking error reduction by exploiting ideas from iterative learning control.
Results on a simulated set-up indicate improved performance over existing identification methods for systems with nonlinearities at the input.
\end{abstract}

\begin{keyword}
Nonlinear system identification, Identification for control, Iterative learning control, Data-based control, Motion Control, Applications in semiconductor manufacturing
\end{keyword}

\end{frontmatter}

\section{Introduction}
\label{chapter:introduction}
The industry of semiconductor manufacturing has ever-increasing demands on manufacturing throughput and accuracy. An example of a manufacturing application is a semiconductor wire bonding machine. In such a machine, a bond head makes interconnections on an integrated circuit, requiring to perform many different motion tasks. These demands are further complicated by nonlinear behavior in the actuators, which is common in many motion system applications. For this reason, increasingly complex control methods are employed that can push the hardware to the limits of its performance.

Advanced feedforward methods can improve tracking performance compared to feedback-only methods. In recent decades, developments in the field of iterative learning control (ILC) 
have enabled data-based methods which can calculate an optimal feedforward signal based on measurements of past experiments. Basis function ILC (BFILC) parameterizes the feedforward signal as a function of the task and learns parameters of a feedforward filter over iterations to obtain a feedforward signal which minimizes the predicted tracking error \citep{VandeWijdeven2010}. BFILC allows for accuracy under varying tasks due to task-dependency of the feedforward signal. The feedforward filter in BFILC typically constitutes an approximate model inverse of the true system \citep{Butterworth2012}. Basis functions can be chosen for instance polynomial \citep{VanderMeulen2008} or can be extended to rational \citep{Blanken2017}. Within rational basis functions, it is possible to pre-specify the locations of the inverse system zeros \citep{Blanken2020}. In \citet{Bolder2015}, an iterative scheme is proposed to instead learn these zeros over iterations. Alternatively, input shaping can be used to compensate system zeros by modifying the task, which results in a convex parameter optimization problem \citep{Bruijnen2012}
. These frameworks with feedforward parametrizations can achieve high tracking accuracy for linear systems while retaining flexibility to task changes.

Importantly, the frameworks mentioned thus far rely on linear parametrizations, which results in limited performance for nonlinear systems. An important class of ILC algorithms called norm-optimal ILC (NOILC) can compensate any repetitive error \citep{D.A.Bristow;M.Tharayil;A.GAlleyne.2006}, even from repetitive nonlinear behavior, as shown in \citet{Gorinevsky2002} and proven in \citet[Chapter 5.3]{Xu2003}. NOILC does this by minimizing the predicted tracking error using a model of the system and signals of past experiments.
However, as it is non-parametric, the high performance is restricted under the assumption of repetitive tasks. Typical semiconductor motion systems operate under varying tasks. For this reason, we look towards methods to model nonlinear behavior parametrically for compensation through feedforward.

A general class of nonlinear systems useful for parametric modeling is called Hammerstein systems \citep{Narendra1966}. These systems consist of a static (memoryless) nonlinear element at the input to a dynamic linear element. Identification of such a system is generally performed by fitting a parameterized model to input/output data obtained from the system \citep{Giri2010}. This model can, for instance, be polynomial in the inputs \citep{Giri2002}, piecewise-linear \citep[Chapter 6]{Giri2010}, or a neural network \citep{Janczak2010}. Non-parametric methods also exist using, for example, regression \citep{Greblicki1989}. The parameters in Hammerstein system identification methods are generally optimized for model accuracy, such that the model prediction error is minimized.

Although many tools exist for data-driven identification of Hammerstein models, these tools all focus on optimizing model accuracy, rather than tracking accuracy of some reference using inverse-model feedforward. At the same time, iterative learning control with basis functions can achieve high tracking accuracy for non-repetitive tasks for linear systems, but fails to compensate for unmodeled nonlinear effects. In contrast, traditional norm-optimal ILC can achieve extreme tracking accuracy even under repetitive nonlinear behavior, but cannot deal with task variations. The aim of this paper is to develop an approach to generate feedforward for Hammerstein systems with high tracking accuracy while retaining task flexibility. The developed approach exploits key ideas from ILC to fit the parameters of a parametric Hammerstein inverse model. These ideas enable the resulting model to have better performance compared to existing Hammerstein identification methods when used for feedforward. Additionally, the proposed approach can be performed in closed-loop, while the optimization can be performed fully off-line.

This paper is structured as follows. In Section \ref{chapter:preliminaries}, preliminary theory relevant to the proposed approach is presented. The problem considered in this paper is presented in Section \ref{chapter:problem}. In Section \ref{chapter:approach}, the proposed approach is introduced. Section \ref{chapter:results} presents the results of the approach applied to a simulated system. Lastly, Section \ref{chapter:conclusions} contains conclusions.

\begin{nota*}
Let $\mathbb{R} =(- \infty,\infty)$. A positive definite matrix $A$ is denoted $A \succ 0$. The weighted 2-norm of a vector $x \in \mathbb{R}^{n }$ with positive definite weighting matrix $W \in \mathbb{R}^{n \times n}$ is denoted by $||x||_W = \sqrt{x^\top W x}$. The $i^{\text{th}}$ element of $x$ is expressed as $x[i]$. The identity matrix of size $n$ is denoted $I_{n}$.

$\boldsymbol{\mathrm{H}}(z)$ denotes a discrete-time (DT), linear time-invariant (LTI), single-input, single-output (SISO) system. Signals are often assumed to be of length $N$. Given input and output vectors $u,y \in \mathbb{R}^{N}$. Let $h(t)$ be the impulse response vector of $\boldsymbol{\mathrm{H}}(z)$. Then, the finite-time response of the possibly noncausal $\boldsymbol{\mathrm{H}}(z)$ to input $u$ is given by the truncated convolution $y[t] = \sum_{l=1-N}^t h(l)u[t-l]$, where $0 \leq t \leq N$ and zero initial and final conditions are assumed, i.e., $u(t) = 0, y(t) = 0$ for all $t < 0$ and $ t \geq N$. The finite-time convolution is denoted as
\vspace{2mm}
\begin{equation*}
\resizebox{.9\hsize}{!}{$
    \underbrace{\begin{bmatrix}
    y[0]\\
    y[1]\\
    \vdots\\
    y[N-1]
    \end{bmatrix}}_y =
    \underbrace{\begin{bmatrix}
    h(0) & h(-1) & \cdots & h(1-N)\\
    h(1) & h(0) & \cdots & h(2-N)\\
    \vdots & \vdots & \ddots & \vdots\\
    h(N-1) & h(N-2) & \cdots & h(0)
    \end{bmatrix}}_H
    \underbrace{\begin{bmatrix}
    u[0]\\
    u[1]\\
    \vdots\\
    u[N-1]
    \end{bmatrix}}_u$},
\end{equation*}
with $H$ the convolution matrix corresponding to $\boldsymbol{\mathrm{H}}(z)$
\end{nota*}

\section{Problem Formulation}
\label{chapter:preliminaries}
In this section, we investigate the problem covered by this paper. Firstly, by introducing and highlighting the limitations of pre-existing frameworks which are nevertheless relevant to the proposed method. Thereafter, the problem is explicitly stated.

\subsection{ILC for repeated tasks}
This section details an important class of ILC algorithms called norm-optimal ILC (NOILC).
\begin{figure}[!t]
    \centering
    \includegraphics[width=60mm]{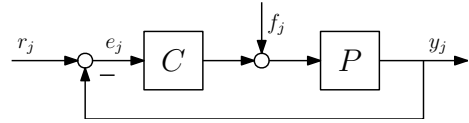}
    \caption{Control scheme with closed-loop feedback and feedforward.}
    \label{fig:control_scheme}   
\end{figure}
Consider the control scheme depicted in Fig.~\ref{fig:control_scheme}. Displayed is a general closed-loop connection with a feedback controller and a feedforward input, which is representative of the application in this paper. In the figure, $r_j \in \mathbb{R}^{N}$ is the reference for iteration or trial $j$, $y_j \in \mathbb{R}^{N}$ is the system output signal, $e_j \in \mathbb{R}^{N} := r_j-y_j$ is the error signal, $u_j \in \mathbb{R}^{N}$ is the system input signal, and $f_j \in \mathbb{R}^{N}$ is the feedforward signal. $C$ represents the DT LTI feedback controller. It is assumed that the closed-loop system is stable. The system $P$ is assumed to be DT, LTI, and SISO. This system is assumed to be rational, i.e.,
\vspace{1mm}
\begin{equation}
\label{eq:rat_plant}
    \boldsymbol{\mathrm{P}}(z) = \frac{\boldsymbol{\mathrm{B}}_0(z)}{\boldsymbol{\mathrm{A}}_0(z)},
\end{equation}

where $\boldsymbol{\mathrm{P}}(z)$ is the corresponding transfer function to the system $P$, and $\boldsymbol{\mathrm{B}}_0(z), \boldsymbol{\mathrm{A}}_0(z)$ are polynomial transfer functions. In order to achieve high tracking accuracy, we want to minimize the next iteration error $e_{j+1}$. To do so, consider expressions for the current and next iteration error signals as a function of the reference and feedforward signal, given by
\vspace{1mm}
\begin{equation}
\label{eq:e_j_general}
    e_j = S r_j - SP f_j,
\end{equation}
\begin{equation}
\label{eq:e_j+1_general}
    e_{j+1} = S r_{j+1} - SP f_{j+1},
\end{equation}

where $S:=(1+P C)^{-1}$ is the sensitivity. We can predict the next iteration error by taking the difference between the current and next iteration error and assuming repetition of the same reference over multiple iterations, i.e., $r_j=r_{j+1}$. This yields the error propagation from iteration $j$ to iteration $j+1$
\vspace{1mm}
\begin{equation}
\label{eq:trial-to-trial_dynamics}
    \hat{e}_{j+1} = e_j + SP (f_j - f_{j+1}),
\end{equation}

where $\hat{e}_{j+1}$ is the predicted error of the next iteration. The objective of ILC is to use data $e_j$ and $f_j$ of the current iteration $j$ to construct the feedforward signal for the next iteration $f_{j+1}$, such that the predicted error of the next iteration $e_{j+1}$ is minimized. This objective can be formulated as an optimization problem with the following cost function
\vspace{1mm}
\begin{equation}
\label{eq:cost_function_general}
    \mathcal{J}(f_{j+1}) := ||\hat{e}_{j+1}(f_{j+1})||_{W_e}^2,
\end{equation}

where $W_e \succ 0$ is a user-defined weighting matrix. This cost function can be extended to include weights on the feedforward signal or changes in the feedforward signal to improve robustness to model uncertainty and iteration-varying disturbances, such as in \citet{D.A.Bristow;M.Tharayil;A.GAlleyne.2006}, or \citet{Gunnarsson2001}. For this paper, the simple form which only penalizes the predicted error suffices.

In general, the minimizer $f^*_{j+1}$ of the cost function can be found by substituting \eqref{eq:trial-to-trial_dynamics} into \eqref{eq:cost_function_general} and solving
\vspace{1mm}
\begin{equation}
\label{eq:cost_function_argmin}
    f^*_{j+1} = \arg \min_{f_{j+1}} \mathcal{J}(f_{j+1}).
\end{equation}

Since the cost function is quadratic in $f_{j+1}$, the optimization problem has a closed-form solution which is obtained by setting the partial derivatives to $f_{j+1}$ to zero. The solution takes the form of an iterative update law for the feedforward signal, see, for instance, \citet{D.A.Bristow;M.Tharayil;A.GAlleyne.2006}.

Note that the presented NOILC scheme converges even for nonlinear systems \citep[Chapter 5.3]{Xu2003}, and can still achieve high tracking accuracy under repetitive nonlinear behavior due to robustness to model uncertainty \citep{Gorinevsky2002}. Additionally, while NOILC is able to achieve very high accuracy for repetitive tasks, i.e., $r_j=r_{j+1}$ \citep{D.A.Bristow;M.Tharayil;A.GAlleyne.2006}, note that the framework is unable to achieve the same accuracy under task variations \citep{Blanken2017}. In order to introduce performance extrapolation to varying tasks, ILC with basis functions is introduced in the next section.

\subsection{ILC for task-flexibility}
\label{subsec:BFILC}
\begin{figure}[t!]
    \centering
    \includegraphics[width=60mm]{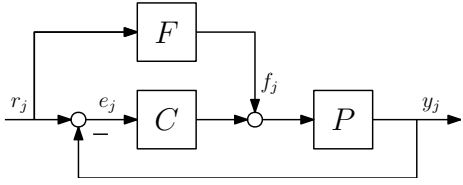}
    \caption{Control scheme with closed-loop feedback and parameterized feedforward.}
    \label{fig:control_scheme_basis_functions}   
\end{figure}
ILC with basis functions (BFILC) is an extension of norm-optimal ILC which enhances its extrapolation capabilities to non-repeating tasks. The key idea in BFILC is that the feedforward signal is now an explicit function of the reference signal, which allows it to adapt to task changes.  Consider the control scheme depicted in Fig.~\ref{fig:control_scheme_basis_functions}. The feedforward signal $f_j$ is constructed by filtering the reference $r_j$ through feedforward filter $\boldsymbol{\mathrm{F}}(z)$. This feedforward filter is in general designed as a parameterized function denoted by
\vspace{1mm}
\begin{equation}
\label{eq:BF_general}
    f_j = F(\theta_j) r_j,
\end{equation}

where $F(\theta_j)$ is the convolution matrix representation of parameterized feedforward filter $\boldsymbol{\mathrm{F}}(\theta_j,z)$, with parameters $\theta_j \in \mathbb{R}^{n_\theta }$. See \citet{Blanken2017} and \citet{VandeWijdeven2010} for similar feedforward structures. By substituting \eqref{eq:BF_general} into \eqref{eq:e_j_general}, we obtain
\vspace{1mm}
\begin{equation}
\label{eq:e_j_BF}
    e_j = S r_j - S P F(\theta_j) r_j.
\end{equation}

From this equation, it can be observed that the reference induced error signal $e_j$ is eliminated for any choice of reference $r_j$ when
\vspace{1mm}
\begin{equation}
\label{eq:bf_objective}
    F = P^{-1}.
\end{equation}

This equation presents the objective for BFILC, in which the parameterized feedforward filter represents an approximate inverse model of the system. Given a parametrized feedforward structure, we seek to optimize the parameters such that they minimize the predicted error $\hat{e}_{j+1}$. This optimization is performed by substitution of \eqref{eq:BF_general} into the previously introduced cost function \eqref{eq:cost_function_general}, and finding the minimizer $\theta^*_{j+1}$ by solving
\vspace{1mm}
\begin{equation}
\label{eq:cost_function_BF_argmin}
    \theta^*_{j+1} = \arg \min_{\theta_{j+1}} \mathcal{J}(\theta_{j+1}).
\end{equation}

The ability of BFILC to perform under task variations comes at the cost of slightly deteriorated performance due to the more restrictive construction of the feedforward signal $f_{j+1}$, which is now constructed from a selection of basis functions.

In BFILC, the choice of basis functions affects the performance due to \eqref{eq:bf_objective}, as well as the properties of the parameter optimization problem. One can choose polynomial basis functions (PBF) which only learn the zero locations of $F(\theta_j)$ resulting in a convex optimization problem with a closed-form solution, see \citet{VandeWijdeven2010}. Alternatively, the feedforward filter $F(\theta_j)$ can be rational with pre-specified pole locations \citep{Blanken2020}. As another option, the pole and zero locations of $F$ can be learned from data, see \citet{Blanken2017}. This results in a non-convex optimization problem that has an iterative solution. Note finally that BFILC, as well as NOILC, can be performed in closed-loop.

While BFILC, either polynomial or rational, can handle task changes, it cannot compensate nonlinear effects. The next section introduces identification tools for a specific class of nonlinear systems called Hammerstein systems.

\subsection{System Identification for Hammerstein Systems}
\begin{figure}[!t]
    \centering
    \includegraphics[width=40mm]{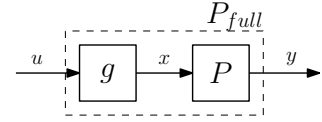}
    \caption{Hammerstein System. Note that the system $P_{full}$ consists of linear subsystem $P$ and static nonlinear subsystem $g$.}
    \label{fig:Hammerstein}   
\end{figure}

Consider Fig.~\ref{fig:Hammerstein}, which shows a general block-oriented system $P_{full}$ consisting of a static nonlinear element $g$ and a linear dynamic element $P$, called a Hammerstein System. In the figure, $u \in \mathbb{R}^{N}$ denotes the system input, $y \in \mathbb{R}^{N}$ denotes the system output, and $x \in \mathbb{R}^{N}$ denotes the intermediate output. To identify the system, one must construct a mapping from inputs $u$ to outputs $y$. To do so, generally, a parametric model $\tilde{g}(\phi)$ of the input nonlinearity $g$ is constructed with parameters $\phi$:
\vspace{1mm}
\begin{equation}
    \hat{x} = \tilde{g}(u,\phi),
\end{equation}

where $\hat{x}$ is the predicted intermediate output, $\tilde{g}(u,\phi)$ is the parameterized input nonlinearity model and $\phi \in \mathbb{R}^{n_\phi }$ are the parameters. Next, a linear discrete-time parametric model is created to represent the linear subsystem $P$ which maps $x$ to $y$ as such:
\vspace{1mm}
\begin{equation}
    \hat{y} = \boldsymbol{\mathrm{\tilde{P}}}(a,b,z) \hat{x},
\end{equation}

where $\hat{y}$ is the predicted output, $\boldsymbol{\mathrm{\tilde{P}}}(a,b,z)$ is the parameterized linear system model which can, for instance, be an infinite impulse response, and $a \in \mathbb{R}^{n_a }$ and $b \in \mathbb{R}^{n_b }$ are the parameters \citep[Chapter 5]{Giri2010}. Note that if the parametric nonlinearity model $\tilde{g}$ is linear in parameters $\phi$, simultaneous optimization of $a$, $b$ and $\phi$ from an input/output dataset using an iterative method is possible \citep{Giri2002}.

Next, we consider how to generate the input/output data, which can be performed by measuring $y$ after applying some input $u$ in open loop. Note that the input signal $u$ must satisfy a persistence of excitation condition for Hammerstein systems \citep{Giri2002}. This can often be satisfied by using white noise which covers the full domain of inputs to be identified in the input nonlinearity.

Lastly, we consider how to optimize the parameter values from the input/output data. The optimization of parameters $\phi$, $a$, $b$ is performed using a dataset consisting of inputs $u$ and matching measured outputs $y$ as the minimizers of a least-squares cost function
\vspace{1mm}
\begin{equation}
\label{eq:argmin_Hammerstein}
\begin{aligned}
    \{a^*,b^*,\phi^*\} = \arg \min_{\{a,b,\phi\}} ||y-\hat{y} ||_2^2,
\end{aligned}
\end{equation}

where $a^*$, $b^*$, and $\phi^*$ are the optimized parameter values \citep[Chapter 5]{Giri2010}. Note that the objective for Hammerstein identification is to minimize the prediction error of the model by matching the parameters to the dataset. This is substantially different from the ILC optimization problem presented in the previous sections, in which the predicted tracking error is minimized.
\label{chapter:problem}
\subsection{Problem Formulation}
The objective of this paper is to identify inverse model feedforward for a Hammerstein system in order to obtain small tracking errors and high flexibility for non-repeating tasks. The theory presented in the previous sections provides useful methods for identification and data-driven compensation. However, currently, no methods exist to satisfy both requirements for the specific class of systems being considered. While methods exist that can estimate parameter values of Hammerstein models, the resulting parameter values are optimized for model accuracy, not small servo error, i.e., error-optimized. And while NOILC can compensate for nonlinear effects, it is inflexible to non-repeating tasks. Lastly, while BFILC can achieve small errors while retaining task flexibility, its performance deteriorates for nonlinear systems. Hence, the formulated problem requires a new method that is parametric, nonlinear, data-driven, and error-optimized.

\section{Approach}
\label{chapter:approach}
\begin{figure}[!t]
    \centering
    \includegraphics[width=80mm]{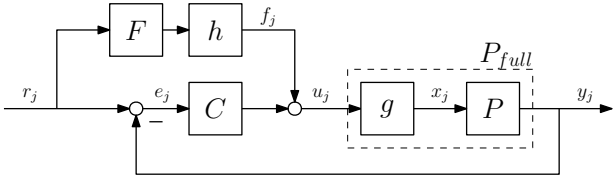}
    \caption{Control scheme with closed-loop feedback and feedforward. Note nonlinear function $h$ in the path of the reference-dependent feedforward signal which aims to compensate the input nonlinearity $g$.}
    \label{fig:solution_control_scheme}   
\end{figure}
The proposed solution is a feedforward controller containing a linear dynamic element $F$ and a static nonlinear element $h$, as shown in Fig.~\ref{fig:solution_control_scheme}. These elements together constitute what is called a Wiener system, as coined in \citet{Schetzen1989}. The key idea in the solution is that we can design $F$ and $h$ such that the error $e_j$ is minimized for any reference $r_j$. Towards this idea, consider the reference-induced error for the control scheme in the figure given by
\vspace{1mm}
\begin{equation}
\label{eq:e_j_nonlin}
    e_j = r_j - P g(C e_j + h(F r_j)).
\end{equation}

We are tasked with eliminating the error $e_j$ by compensation of the linear and nonlinear subsystems. Zero reference-induced error, i.e., $e_j = 0$ for any reference $r_j$ in \eqref{eq:e_j_nonlin} is achieved for
\vspace{1.5mm}
\begin{equation}
    P g(h(F r_j)) = r_j,
\end{equation}

which is satisfied if
\vspace{1.5mm}
\begin{equation}
\label{eq:nonlin_inversion_obj}
\begin{aligned}
    F = P^{-1},\\
    h = g^{-1}.
\end{aligned}
\end{equation}

Recall \eqref{eq:bf_objective}, which presents the same objective as the first equation here. Hence, a parametrization for $F$ according to BFILC is sufficient to allow for inversion of $P$. The second objective can be satisfied by an inverse model $h$ of the nonlinearity $g$. $h$ can be parametrized using existing methods for static nonlinearity modeling, such as polynomial, piecewise linear, etc.

Next, we consider how to generate the input/output data for the parameter optimization problem. The key idea here is that we can curate the dataset to be more relevant to the minimization of the tracking error. Consider that $F$ and $h$ are typically under-modeled due to high-order dynamics, damped flexible modes, and complex unknown input nonlinearities. Therefore, the optimal parameter values in $F$ and $h$ are always a compromise. However, some system behavior is more relevant for typical tasks performed by the system. Recall that after converging, NOILC finds a signal which inverts the repetitive system behavior. Hence, this converged signal contains information about the task-relevant system behavior. Fitting to this optimal feedforward signal swings the parameter compromise in the favor of error reduction. For this reason, we use NOILC to generate a reference-feedforward mapping to which the parameters of $F$ and $h$ are fitted.

We train the NOILC feedforward signal with a combined setpoint that contains multiple typical references in order to reduce bias in the dataset towards a single reference. Note that persistence of excitation, in this case, is satisfied if the reference used requires a feedforward signal which covers the full domain of inputs that is typically used for the application. In practice, this can be satisfied by using a smooth and challenging reference. Note furthermore that the generation of data can be performed in closed-loop using this method.

Lastly, we consider how the parameter values are optimized from the dataset. The parameter optimization problem can be formulated through the cost function
\vspace{1.5mm}
\begin{equation}
\label{eq:cost_function_nonlin}
    \mathcal{K}(\theta, \phi) = ||SP(f_{NOILC}-h(F(\theta)r,\phi)) ||_{W}^2,
\end{equation}

where $f_{NOILC}$ is the converged NOILC feedforward signal corresponding to combined reference $r$, $\theta$ and $\phi$ are the parameters, and $W$ is a user-defined weighting matrix. In the cost function, another key idea from ILC is employed. The feedforward signal error is filtered by the process sensitivity in order to weigh the feedforward difference by contribution to the resulting error, similar to \eqref{eq:trial-to-trial_dynamics} in \eqref{eq:cost_function_general} and as seen in \citet{Aarnoudse2021}. The inclusion of this filter is what makes the optimized parameter values performance-relevant. We can find the optimal parameter values $\theta^*$ and $\phi^*$ by solving:
\vspace{1.5mm}
\begin{equation}
\label{eq:cost_function_nonlin_argmin}
    \{\theta^*, \phi^*\} = \arg \min_{\{\theta, \phi\}} \mathcal{K}(\theta, \phi).
\end{equation}

Note that if $h$ is linear in parameters $\phi$, an iterative solution to the parameter optimization problem is possible, similar to existing Hammerstein system identification methods. If additionally $F$ is linear in parameters $\theta$, the parameter optimization problem is convex and has a closed-form solution. If this is not the case, one must rely on non-convex optimization solvers and a global optimum cannot be guaranteed.

This section presented a parametrized Hammerstein system identification method that will result in a performance-relevant fit and which uses a more performance-relevant dataset. In the next section, we will evaluate the effectiveness of the proposed method on a simulated setup.

\section{Results on Simulated System}
\label{chapter:results}
In this section, the performance of the proposed approach is compared against pre-existing approaches on a simulation of a wire bonder by ASM Pacific Technology (PT), a leading manufacturer of semiconductor equipment.

\subsection{Simulated Set-up}
\label{subsec:sim_sys}
The simulated system is a multibody model of a real ASM PT wire bonder in Simscape. The motion system model considers multiple rigid bodies, connected by stiffnesses and dampers. Only the SISO x-direction is considered. The linear system is rational, i.e., see \eqref{eq:rat_plant}. Furthermore, the system contains an input nonlinearity in the form of magnetic saturation. Magnetic saturation causes a force drop-off as the input current on the motor increases. This phenomenon is modeled through the function
\vspace{1.5mm}
\begin{equation}
\label{eq:magnetic_saturation}
    g(u_j) = I_{max} \cdot \mathrm{tanh}(u_j/ I_{max}),
\end{equation}

where $u_j$ is the input signal in amperes [A], and $I_{max} \in \mathbb{R}_{>0}$ is the constant saturation parameter in amperes [A] \citep{Na2018}. A saturation parameter value of $I_{max} = 70$ [A] is used in the simulations.

\begin{figure}[!t]
    \centering
    \includegraphics[width=75mm]{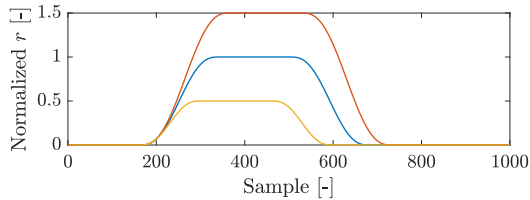}
    \caption{Normalized references $r^1$ ({\color[rgb]{0.00,0.45,0.74}-}), performed in the first 3 iterations; $r^2$ ({\color[rgb]{0.85,0.33,0.1} -}), performed in the next 2 iterations; and $r^3$ ({\color[rgb]{0.93,0.69,0.13} -}), performed in the last 2 iterations.}
    \label{fig:references}   
\end{figure}

In order to examine performance extrapolation under varying tasks, multiple references are used. Fig.~\ref{fig:references} shows the set of references used on the simulated system. The references are all quintic polynomial trajectories \cite[Section 5.6]{Spong2005}. Note that these references have different motion distances and maximum accelerations.

Lastly, the system is subjected to Gaussian noise at the output measurement with a variance of $7.5 \cdot 10^{-6}$ times smaller than the motion distance of $r^1$. The next section details the application of the developed approach for the simulated setup.

\subsection{Identification of Hammerstein system for feedforward control}
\label{subsec:sim_approach}
The proposed approach consists of modeling, generation of input/output data, and parameter optimization. We first model $F(\theta)$ as PBF rigid-body feedforward with velocity and acceleration bases, i.e.,
\vspace{1.5mm}
\begin{equation}
\label{eq:F_model}
    \boldsymbol{\mathrm{F}}(\theta,z) = \left(1-z^{-1}\right) \theta[1] + \left(1-z^{-1}\right)^2 \theta[2].
\end{equation}

Note that the model is static in this case, i.e., $\theta_j = \theta, \forall j$.

Next, we model the inverse input nonlinearity. The parameterized nonlinear function $h$, which is chosen as the inverse of the magnetic saturation model in \eqref{eq:magnetic_saturation}, is given by
\vspace{1.5mm}
\begin{equation}
\label{eq:h_model}
    h(F(\theta)r,\phi) = \phi \cdot \mathrm{atanh}(F(\theta)r / \phi),
\end{equation}

where $r$ is the combined reference used for identification, $\phi$ is the nonlinearity parameter in [A], which models the saturation parameter $I_{max}$ in \eqref{eq:magnetic_saturation}.

\begin{figure}[!t]
    \centering
    \includegraphics[width=80mm]{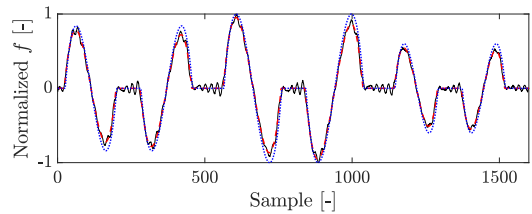}
    \caption{Normalized converged NOILC feedforward $f_{NOILC}$ signal ({\color{black} -}) for combined reference $r$, fitted nonlinearity filtered rigid-body feedforward signal ({\color{red} -{}-}) and scaled acceleration of combined reference $r$ ({\color{blue} \raisebox{1.3pt}{...}}).}
    \label{fig:nonlinearity_feedforward_fit}   
\end{figure}

Next, we detail the generation of input/output data. A combined setpoint is used which contains a sequence of 3 quintic polynomial reference profiles with varying distance and maximum acceleration. The NOILC algorithm is applied to this reference for 10 iterations to obtain $f_{NOILC}$. As seen in Fig.~\ref{fig:nonlinearity_feedforward_fit}, $f_{NOILC}$ resembles the shape of the acceleration of the reference $r$, as the system is dominated by rigid-body dynamics. Hence, by choosing a smooth and challenging reference, the resulting optimized NOILC feedforward signal will contain the full domain of possible typical inputs.

Lastly, we optimize the parameters $\theta$ and $\phi$ using cost function \eqref{eq:cost_function_nonlin}. Non-convex optimization is employed, as $\phi$ appears nonlinearly in $h$, see \eqref{eq:h_model}. In this paper, an unconstrained particle swarm optimization is used with a swarm size of 200 \citep{Kennedy1995}. Furthermore, the cost function weight is chosen as $W=I_{N}$. This results in the fit shown in Fig.~\ref{fig:nonlinearity_feedforward_fit}.The optimized parameter value converges to $\phi^* = 69.6$ [A], which closely approximates saturation parameter $I_{max} = 70$ [A] used in the simulation.

In the next section, we evaluate how the identified parametric Hammerstein model performs in terms of tracking accuracy and flexibility to task changes in comparison to the pre-existing methods.

\subsection{Performance comparison}
We compare the proposed approach to linear BFILC with velocity and acceleration bases, i.e., $h(F r_j) = F r_j$. Note that in this approach, the parameters are retuned after each iteration $j$ based on data from the previous iteration.

Additionally, the proposed approach is compared against the performance of inverse model feedforward using existing Hammerstein system identification approaches. This approach uses \eqref{eq:F_model} as $\tilde{P}^{-1}$ and \eqref{eq:h_model} as $\tilde{g}^{-1}$. The parameters of these models are optimized using particle swarm as detailed in Section \ref{subsec:sim_approach}, except with white noise open-loop input/output data and parameter optimization problem \eqref{eq:argmin_Hammerstein}. This method optimizes to $\phi^* = 163.9$, which is significantly different from the true value of $I_{max}$.

\begin{figure}[!t]
    \centering
    \includegraphics[width=80mm]{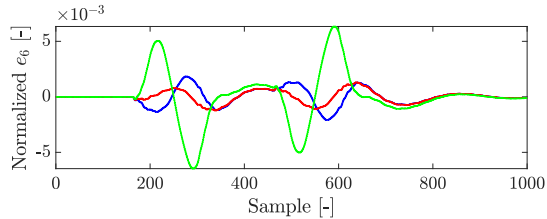}
    \caption{Error signal of trial 6 for linear parameterized feedforward ({\color{blue}-}), inverse model feedforward using existing Hammerstein identification methods \mbox{({\color{green}-})}, and the proposed method ({\color{red}-}), normalized to the motion distance of $r^1$. Note that this trial occurs directly after a task change.}
    \label{fig:BF_vs_proposed}   
\end{figure}

\begin{figure}[!t]
    \centering
    \includegraphics[width=80mm]{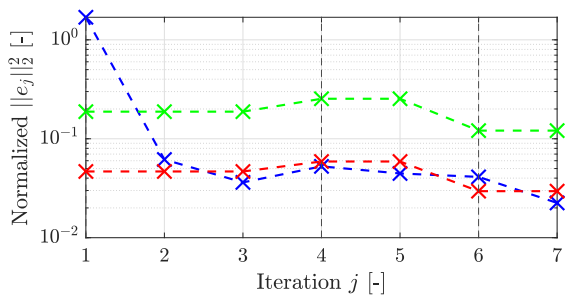}
    \caption{Comparison of the normalized error 2-norm per trial for linear parameterized feedforward ({\color{blue} -}), inverse model feedforward using existing Hammerstein identification methods ({\color{green} -}), and the proposed method ({\color{red} -}). To show the flexibility of the methods, the motion task is changed at trials 4 and 6, which is indicated by vertical dashed lines in the figure.}
    \label{fig:error_2_norm_comparison}   
\end{figure}

The results of the comparison are shown in Fig.~\ref{fig:BF_vs_proposed} and Fig.~\ref{fig:error_2_norm_comparison}.

Observe from Fig.~\ref{fig:BF_vs_proposed} that the proposed method has lower tracking error than both linear BFILC and inverse model feedforward using existing Hammerstein identification methods on the first trial of the third reference. Note that the remaining error for both BFILC and the proposed method oscillates, due to the presence of a flexible mode that cannot be compensated by rigid-body feedforward.

Fig.~\ref{fig:error_2_norm_comparison} shows a comparison between the methods in terms of error 2-norm per trial. The proposed method results in a model that generates better feedforward than traditional Hammerstein system identification methods for all references. Observe that the proposed method has a lower error 2-norm directly after the second task change compared to linear parameterized feedforward, due to compensation of the input nonlinearity. Note that linear BFILC is able to learn from repetitions of the reference, yielding a lower error 2-norm by overfitting the current task, but is unable to compensate for nonlinear effects.

\section{Conclusion}
\label{chapter:conclusions}
This paper introduces a method for identification and compensation of Hammerstein systems by exploiting ideas from iterative learning control. This contrasts with existing Hammerstein identification methods by making the parameter estimation problem task-relevant and performance-relevant. The results on a simulated motion system indicate that the method is promising in terms of reducing tracking error under varying tasks.

Future research is towards simultaneous learning of linear and nonlinear parameters within the iterative learning control framework and analysis of the parameter optimization problem.

\section{Acknowledgment}
The authors would like to thank Robin van Es for his contributions to this research.

{
\footnotesize
\setlength{\bibsep}{0pt plus 0.3ex}
\bibliography{ifacconf}
}
\end{document}